\documentclass[twocolumn,superscriptaddress,showpacs,aps,prl,longbibliography]{revtex4-2}
\usepackage[utf8]{inputenc}
\usepackage[english]{babel}
\usepackage[T1]{fontenc}
\usepackage{microtype}
\usepackage[pdftex]{graphicx}
\usepackage{hyperref}
\usepackage{amsmath}
\usepackage{amssymb}
\usepackage{braket}
\usepackage{xcolor}
\usepackage{dcolumn}
\usepackage{bbm}
\usepackage{bbold}
\usepackage[caption=false]{subfig}
\usepackage[capitalize]{cleveref}
\usepackage{float}
\usepackage{placeins}

\usepackage{soul,xcolor}
\usepackage[normalem]{ulem}
\definecolor{black}{rgb}{0,0,0}
\definecolor{blue}{rgb}{0,0,1}
\definecolor{green}{rgb}{0,1,0}
\definecolor{red}{rgb}{1,0,0}
\definecolor{brown}{rgb}{0.4,0.2,0}
\definecolor{darkgreen}{rgb}{0,0.7,0}
\definecolor{darkblue}{rgb}{0.0,0.0,0.5}
\definecolor{red}{rgb}{1,0,0}
\definecolor{deepmagenta}{rgb}{0.8, 0.0, 0.8}

\hypersetup{
    hypertexnames=false,
    colorlinks=true,
    allcolors=blue,
    unicode=true
}

\newcommand{\phantomsubfloat}[1]{
    {
        \captionsetup[subfloat]{farskip=0pt,captionskip=0pt}
        \captionsetup[subfigure]{labelformat=empty}
        \subfloat{#1}
    }
}

\graphicspath{{./Figures/}}
\usepackage{titlesec}

\def \IBK{Institute for Theoretical Physics, University of Innsbruck, Innsbruck 6020, Austria}
\def \IQOQI{Institute for Quantum Optics and Quantum Information,
	Austrian Academy of Sciences, Innsbruck 6020, Austria}
\def \JILA{JILA, University of Colorado and National Institute of Standards and Technology,
and Department of Physics, University of Colorado, Boulder, Colorado 80309, USA}

\begin{document}

\title{Adiabatic echo protocols for robust quantum many-body state preparation}
	\author{Zhongda Zeng}\thanks{These authors contributed equally to this work.}\affiliation{\IBK}\affiliation{\IQOQI}
	\author{Giuliano Giudici}\thanks{These authors contributed equally to this work.}\affiliation{\IBK}\affiliation{\IQOQI}
    \author{Aruku Senoo}\affiliation{\JILA}
    \author{Alexander~Baumgärtner}\affiliation{\JILA}
    \author{Adam M. Kaufman}\affiliation{\JILA}
	\author{Hannes Pichler}\email{hannes.pichler@uibk.ac.at}\affiliation{\IBK}\affiliation{\IQOQI}
\preprint{}

\begin{abstract}
Entangled many-body states are a key resource for quantum technologies. Yet their preparation through analog control of interacting quantum systems is often hindered by experimental imperfections. Here, we introduce the adiabatic echo protocol, a general approach to state preparation designed to suppress the effect of static perturbations. We provide an analytical understanding of its robustness in terms of dynamically engineered destructive interference. By applying quantum optimal control methods, we demonstrate that such a protocol emerges naturally in a variety of settings, without requiring assumptions on the form of the control fields. Examples include Greenberger-Horne-Zeilinger state preparation in Ising spin chains and two-dimensional Rydberg atom arrays, as well as the generation of quantum spin liquid states in frustrated Rydberg lattices. Our results highlight the broad applicability of this protocol, providing a practical framework for reliable many-body state preparation in present-day quantum platforms.
\end{abstract}

\date{\today}

\maketitle

\paragraph{Introduction. --}
High-fidelity preparation of many-body quantum states is essential for advancing quantum technologies, with applications ranging from quantum computation and simulation~\cite{NielsenChuang2000,Farhi2000,Raussendorf2001,Cirac2012,Georgescu2014,Gross2017,Altman2021} to precision sensing~\cite{Giovannetti2011,Degen2017,Ye_2024}. 
Analog approaches provide a natural and well-established route via continuous steering of the system through time-dependent control fields~\cite{Dziarmaga2010,Polkovnikov2011,Kolodrubetz2017,Sels2017,Albash_2018}. 
However, their effectiveness is often limited by experimental imperfections, such as static perturbations that alter the system dynamics in an uncontrolled fashion, thereby reducing the fidelity with the target state~\cite{Hauke2012}. This motivates the development of control protocols that not only enable accurate state preparation in an ideal setting, but also ensure robustness against such imperfections under realistic conditions.

\begin{figure}[t!]
    \phantomsubfloat{\label{fig1:a}}
    \phantomsubfloat{\label{fig1:b}}
    \phantomsubfloat{\label{fig1:c}}
    \phantomsubfloat{\label{fig1:d}}
    \phantomsubfloat{\label{fig1:e}}
    \centering  
    \includegraphics[width=1.0\columnwidth]{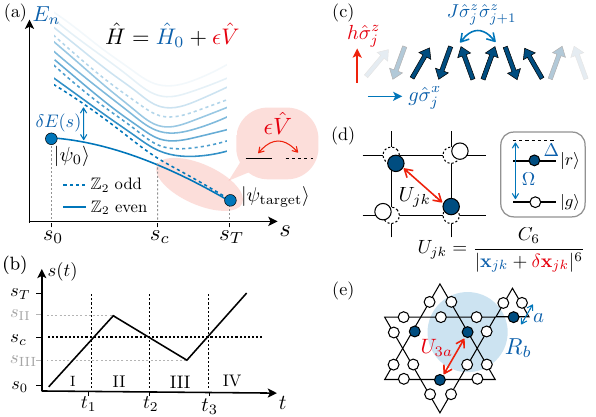}
    \vspace*{-2mm}
    \caption{\textbf{Many-body adiabatic echo protocol.} (a)~Spectrum of a many-body Hamiltonian $\hat{H}_0(s)$ with $\mathbb{Z}_2$ symmetry breaking. The initial and target states lie in the trivial and ordered phases, separated by a critical point $s_c$. Solid (dashed) lines indicate eigenenergies in the even (odd) $\mathbb{Z}_2$ sectors. A perturbation $\hat{V}$ breaks the symmetry and couples these sectors, affecting the dynamics for $s > s_c$.
    (b)~Illustration of the adiabatic echo protocol. Contributions to the infidelity $\mathcal{I}$ from regions I and III vanish, while those from regions II and IV destructively interfere.  (c–e)~Applications where the echo protocol mitigates the effect of the perturbation $\hat{V}$: (c)~Ising chain with transverse and longitudinal fields, (d)~square lattice Rydberg array with positional disorder, (e)~ruby lattice Rydberg array with long-range interactions. }
    \label{fig1}
\end{figure}

Given a time-dependent Hamiltonian $\hat{H}_0(s(t))$, where $s(t)$ is a control parameter, we focus on the problem of preparing a strongly-correlated ground state at $s(T) = s_T$ by evolving the system from a trivial ground state at $s(0) = s_0$.
The buildup of quantum correlations in the target state typically requires crossing a continuous phase transition, a paradigm at the core of recent analog quantum simulation experiments~\cite{Bernien2017,Keesling2019,Omran2019,Ebadi_2021,Scholl2021,Semeghini2021,Guo2024}.
Our goal is to devise control protocols that are robust against static perturbations $\epsilon \hat{V}$ (cf.~\cref{fig1}), i.e., insensitive to small variations in $\epsilon$, which is fixed during the evolution but may vary across experimental runs.

To this end, we introduce a generic preparation strategy that emerges from optimal control~\cite{Glaser2015} and can be implemented across a broad class of many-body Hamiltonians.  
We dub this protocol \emph{adiabatic echo protocol}, as it operates by dynamically “echoing out” the effect of static perturbations during an adiabatic evolution.
After outlining the general framework in which the protocol is applicable, we develop an analytical description of the mechanism behind its robustness and validate it numerically in a concrete many-body setting.
We then show how it naturally arises in several interacting models using the Gradient Ascent Pulse Engineering (GRAPE) algorithm~\cite{Khaneja2005}--a method widely used in few-qubit gate design~\cite{Garon2013,Egger2014,Dolde2014,Jandura2022,Giudici2024} but rarely applied in the context of many-body state preparation.
In particular, we showcase its versatility in various contexts: the preparation of Greenberger-Horne-Zeilinger (GHZ) states via crossing a symmetry-breaking phase transition in the quantum Ising chain~\cite{Wang2010} and in square Rydberg lattices~\cite{Omran2019}, and the generation of quantum spin liquid (QSL) states via crossing a topological phase transition in frustrated Rydberg arrays~\cite{Semeghini2021}.
We further confirm the benefits of the adiabatic echo protocol in a companion experimental work by preparing GHZ states in Rydberg atom ladders~\cite{Senoo2025}. 

\paragraph{Adiabatic echo protocol. --} 

Dynamical preparation of many-body states often relies on the adiabatic theorem, which ensures that a system governed by a Hamiltonian $\hat{H}_0(s(t))$ remains close to its instantaneous ground state when $s(t)$ is varied slowly in time, relative to the inverse energy gap~\cite{Albash_2018}. 
Therefore, driving the system from an easy-to-prepare ground state $\ket{\psi_0}$ at $s_0$ to an entangled target ground state $\ket{\psi_{\mathrm{target}}}$ at $s_T$ through a quantum phase transition, where energy gaps close polynomially with system size $L$, requires preparation times that scale algebraically with $L$ to achieve a fixed fidelity~\cite{Albash_2018}.

In the simplest scenario, the phase transition occurs via spontaneous symmetry breaking, with a critical point $s_c$ separating a trivial phase ($s < s_c$) from an ordered phase ($s > s_c$).  
The initial and final values of the control function, $s_0$ and $s_T$, lie in the trivial and ordered phases, respectively.  
Below, we focus on $\mathbb{Z}_2$ symmetry breaking and refer to \cite{SupMat_arxiv} for a discussion of higher-order symmetry groups.  
The typical energy spectrum of a many-body Hamiltonian undergoing such a transition is sketched in \cref{fig1:a}: for $s < s_c$, the ground state $\ket{E_0(s)}$ is well separated from the excited states, while for $s > s_c$, spontaneous symmetry breaking leads to an exponentially small gap between the ground state and the first excited state $\ket{E_1(s)}$.  
We assume, without loss of generality, that these two states belong to different $\mathbb{Z}_2$ symmetry sectors and that no level crossing occurs for any value of $s$ \footnote{Level crossings between the two symmetry sectors occurring at isolated points would lead only to subleading contributions to the preparation infidelity induced by symmetry breaking. As shown in \cite{SupMat_arxiv}, the dominant contribution instead arises from the ordered phase, where hybridization between the ground and first excited states persists over a time interval that scales extensively with system size.}.

In the absence of perturbations, symmetry forbids mixing between $\ket{E_0(s)}$ and $\ket{E_1(s)}$. However, static symmetry-breaking terms $\epsilon \hat{V}$--arising from experimental imperfections or theoretical approximations--induce a nonzero coupling $V_{10}(s) = \bra{E_1(s)} \hat{V} \ket{E_0(s)}$.
In the adiabatic limit, couplings to other higher excited states can be neglected, and the leading-order contribution to the preparation infidelity $\mathcal{I} = 1 - \left| \langle \psi_{\mathrm{target}} | \psi(T) \rangle \right|^2$ is given by time-dependent perturbation theory as~\cite{SupMat_arxiv}
\begin{equation}
\mathcal{I} = \epsilon^2 \left| \int_{0}^{T} \! dt \, V_{10}(s(t)) \, e^{-i \int_{t}^{T} \! d\tau \, \delta E(s(\tau))} \right|^2 ,
\label{eq:F}
\end{equation}
where $\delta E(s) = E_1(s) - E_0(s)$ is the lowest energy gap.

The dominant contributions to the infidelity generically arise from the region where $\delta E(s)$ is exponentially small with the system size $L$ ($s > s_c$), while contributions from the finite-gap region ($s < s_c$) are subleading and parametrically suppressed with $L$. We neglect the latter contributions in the following and refer to~\cite{SupMat_arxiv} for a formal justification of this approximation.

In a standard adiabatic protocol, $s(t)$ is monotonic and the critical point $s_c$ is crossed only once at $t = t_c$. Because of the exponentially small gap, the phase of the integrand in \cref{eq:F} simplifies, and the infidelity reduces to $\mathcal{I} \simeq \epsilon^2 |V_{10}|^2 (T - t_c)^2$, where we assumed $V_{10}(s) \simeq V_{10}$.
Hence, while increasing the total evolution time $T$ improves adiabaticity, it amplifies the infidelity in the presence of symmetry-breaking perturbations.  

This trade-off can be avoided by protocols that sweep across the phase transition multiple times, such as the one in \cref{fig1:b}, which consists of four segments. In segments I and III, $s < s_c$, while in II and IV, $s > s_c$.  
Neglecting the contribution to the infidelity for $s < s_c$, we have 
\begin{equation}
    \mathcal{I} \simeq \epsilon^2 \left| \, e^{-i\alpha}  \int_{t_1}^{t_2} dt \, V_{10}(s(t)) + \int_{t_3}^{T} dt \, V_{10}(s(t)) \, \right|^2,
\end{equation}
where $\alpha = \int_{t_2}^{t_3} d\tau\, \delta E(s(\tau))$.  
By designing the protocol $s(t)$ such that $\alpha = \pi$ and $\int_{t_1}^{t_2} dt \, V_{10} = \int_{t_3}^{T} dt \, V_{10}$, these two amplitudes interfere destructively, thereby eliminating preparation infidelities at leading order in $\epsilon$. We define a protocol satisfying these two conditions as an \emph{adiabatic echo protocol}, where the name stems from its analogy with spin echo sequences~\cite{spinecho}. While in that context dynamical decoupling cancels coherent errors at the Hamiltonian level~\cite{Ernst1990}, in our many-body setting the cancellation occurs in the instantaneous adiabatic eigenbasis.

The analytical argument above identifies a family of control pulses with a universal structure that ensures robustness against static perturbations but does not uniquely fix the control profile. In what follows, we consider concrete model Hamiltonians for the preparation of selected many-body target states and show how adiabatic echo protocols can be designed in practice. We first demonstrate that, within a simple manifold spanned by two parameters that set the relative phase and amplitude for destructive interference, the adiabatic echo protocol coincides with the optimal robust solution. We then use GRAPE to show how adiabatic echo protocols emerge in unbiased optimal control across multiple interacting models and target states. We further show in~\cite{SupMat_arxiv} that these results are independent of the optimal control method employed.

\begin{figure*}[t]
    \phantomsubfloat{\label{fig2:a}}
    \phantomsubfloat{\label{fig2:b}}
    \phantomsubfloat{\label{fig2:c}}
    \phantomsubfloat{\label{fig2:d}}
    \phantomsubfloat{\label{fig2:e}}
    \phantomsubfloat{\label{fig2:f}}
    \centering  
    \includegraphics[width=1\linewidth]{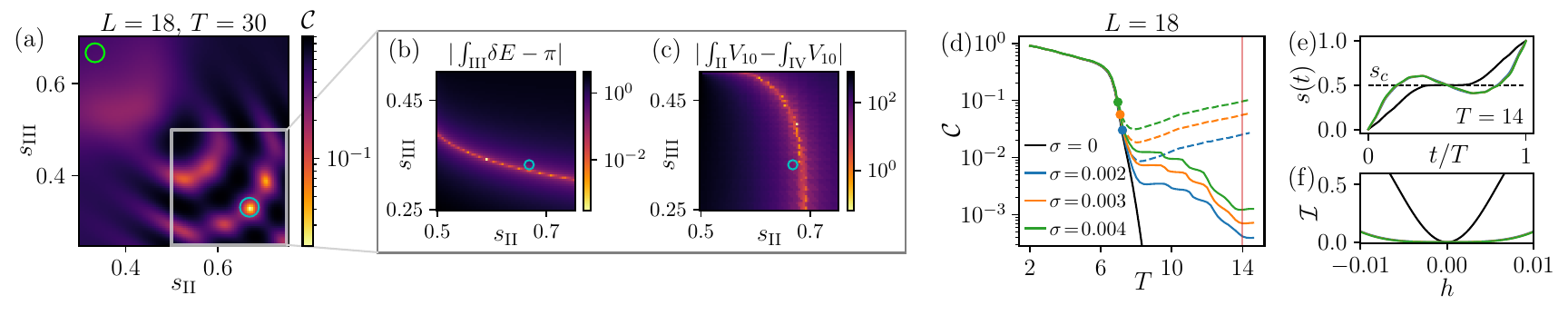}
    \vspace*{-5mm}
    \caption{\textbf{Adiabatic echo in the quantum Ising chain.} (a)~Average infidelity \cref{eq:cost_f} for the protocol in \cref{fig1:b} in the $(s_{\rm II},s_{\rm III})$ plane, for $\sigma=0.003$, $N_{\rm s}=30$, and $T=30$. The light blue circle indicates the optimal solution, while the green circle denotes the standard linear ramp. (b--c)~Interference conditions defining the adiabatic echo protocol in the $(s_{\rm II},s_{\rm III})$ plane; the minimum in (a) (light blue circle) lies close to their intersection. (d)~Optimal cost function $\mathcal{C}$ versus total preparation time $T$ for different $\sigma$ ($N_{\rm s}=30$). Dashed lines show the averaged infidelity evaluated on time-optimal protocols obtained for $\sigma=0$. Colored dots indicate the time $T^*$ beyond which the $\sigma>0$ optimization deviates from the time-optimal solution. (e)~Optimized control profiles $s(t)$ for $T=14$ (vertical red line in (d)) and different $\sigma$. (f)~Preparation infidelity for the profiles in (e) as a function of the longitudinal field $h$.}
    \label{fig2}
\end{figure*}

\paragraph{GHZ state preparation in the quantum Ising chain. --} 
The quantum Ising chain Hamiltonian reads~\cite{Suzuki_2013}
\begin{equation}
\hat{H}_{\rm Ising} = - J\sum_j \hat{\sigma}_{j}^{z} \hat{\sigma}_{j+1}^{z} - g \sum_j \hat{\sigma}_j^x + h \sum_j \hat{\sigma}_j^z \, ,
\label{eq:ising}
\end{equation}
where $\hat{\sigma}_j^{x/z}$ are Pauli matrices and periodic boundary conditions are assumed. 
We parametrize the couplings $J$ and $g$ with a single control parameter $s$ as $J=J_0 \sin(\pi s / 2)$ and $g= J_0 \cos(\pi s/2)$, and fix the overall energy scale by setting $J_0 = 1 $.
For $h=0$, $\hat{H}_{\rm Ising}$ represents the unperturbed Hamiltonian $\hat{H}_0$, which has a $\mathbb{Z}_2$ symmetry generated by $\prod_j \hat{\sigma}^x_j$. The critical point $s_c = 1/2$ separates the disordered phase ($s<s_c$) from the ordered one ($s>s_c$). The last term in \cref{eq:ising} breaks the symmetry and plays the role of the static perturbation $\hat{V}$.

The dynamical preparation is carried out by evolving the system under a time-dependent $s(t)$, starting from the product state $\ket{+ +\cdots +}$, i.e., the ground state of $\hat{H}_0(s=0)$. 
The target state is $\ket{ \psi_{\rm GHZ}^+ } =\frac{1}{\sqrt{2}}(\ket{ \uparrow \uparrow \cdots\uparrow} +\ket{ \downarrow \downarrow \cdots\downarrow})$, i.e., the ground state of $\hat{H}_0(s=1)$, exactly degenerate with $\ket{ \psi_{\rm GHZ}^- } =\frac{1}{\sqrt{2}}(\ket{ \uparrow \uparrow \cdots\uparrow} - \ket{ \downarrow \downarrow \cdots\downarrow})$. Both the initial and target states belong to the even $\mathbb{Z}_2$ symmetry sector. The static field $h$ couples these two sectors, inducing leakage from $\ket{ \psi_{\rm GHZ}^+ }$ to $\ket{ \psi_{\rm GHZ}^- }$ when $s(t) \gtrsim s_c$, resulting in a loss of fidelity for the prepared GHZ state.

We begin by considering a preparation protocol of the form depicted in \cref{fig1:b}, i.e. a piecewise linear $s(t)$ satisfying $s(0) = 0$, $s(T) = 1$, $s(T/3) = s_{\rm II}$, and $s(2T/3) = s_{\rm III}$, where $s_{\rm II}$ and $s_{\rm III}$ are two control parameters. To quantify robustness against symmetry breaking, we use the preparation infidelity averaged over $N_{\rm s}$ values of the perturbation strength $\epsilon_\ell$:
\begin{equation}
    \mathcal{C} = 1 - \frac{1}{N_{\rm s}} \sum_{\ell=1}^{N_{\rm s}} \left| \braket{ \psi_{\rm target}  | \! \prod_{j=0}^{N-1} \! e^{-i dt \left[ \hat{H}_0(s_j) + \epsilon_\ell \hat{V} \right] } | \psi_{0} } \right|^2,
    \label{eq:cost_f}
\end{equation}
where $s_j = s(t_j)$, $t_j = ( j + \frac{1}{2} ) d t $ with $j = 0,\dots N-1$, $dt = T/N$ with $T$ the total evolution time, and $\epsilon$ is the longitudinal field $h$ in \cref{eq:ising}, sampled uniformly in the interval $[-\sigma,\sigma]$.
The case $\sigma = 0 $ ($N_{\rm s} = 1$) corresponds to the unperturbed control problem.

In \cref{fig2:a} we plot $\mathcal{C}$ in the $(s_{\rm II},s_{\rm III})$ plane for $L=18$ spins and $T=30$, revealing a clear global minimum for $(s_{\rm II},s_{\rm III}) \simeq (0.67,0.33)$. Remarkably, this point lies at the intersection of the two interference conditions defining the adiabatic echo protocol: the accumulation of a dynamical phase equal to $\pi$ in region $\rm III$ (\cref{fig2:b}) and equal symmetry-breaking transition amplitudes in regions $\rm II$ and $\rm IV$ (\cref{fig2:c}).

We now apply GRAPE to optimize the function $s(t)$ in the full control space for preparing $\ket{ \psi_{\rm GHZ}^+ }$. The variational parameters in this case are the $N$ time-discretized values $s_j$. The optimization is performed using standard routines that rely on the gradient $\partial C/\partial s_j$, which can be efficiently computed numerically~\cite{SupMat_arxiv}. Moreover, we add a term to $\mathcal{C}$ that penalizes sharp variations of the optimal control function between consecutive time steps~\cite{SupMat_arxiv}.
In \cref{fig2:d}, we plot the optimal $\mathcal{C}$ as a function of the preparation time $T$ for various values of the maximal perturbation strength $\sigma$, for $L=18$. The system-size scaling of these results is discussed in \cite{SupMat_arxiv}. The solid black line represents purely time-optimal protocols obtained for $\sigma=0$,  while the dashed lines represent the corresponding averaged infidelities \cref{eq:cost_f} evaluated for $\sigma > 0$.

For small $T$, the infidelity is dominated by non-adiabatic transitions within the even $\mathbb{Z}_2$ sector, and the optimization for $\sigma > 0$ leads to the same results as for $\sigma = 0$. 
As $T$ increases and the evolution turns adiabatic, the leakage to the odd $\mathbb{Z}_2$ sector due to the perturbation becomes the main source of infidelity. In this regime, the optimal solution for $\sigma > 0$ deviates from the $\sigma = 0$ case and takes the form of an echo protocol. This is illustrated in \cref{fig2:e}, where a typical time-optimal solution (black line), in which $s(t)$ crosses $s_c$ only once, is compared with an echo-like solution which resembles a smoothed version of \cref{fig1:b}. Notably, the optimal control solution is effective at shorter preparation times than those considered in \cref{fig2:a}.

Although the evolution time $T^*$ at which the echo protocol becomes beneficial decreases as $\sigma$ increases (cf. the colored markers in \cref{fig2:d}), the control profile $s(t)$ appears to be insensitive to the value of $\sigma$. 
In \cref{fig2:f}, we plot the infidelity as a function of the longitudinal field $h$ for the echo protocol (colored lines), demonstrating its significantly enhanced robustness compared to the time-optimal case (black line). We stress that this property is not limited to the particular perturbation considered here but extends to more general symmetry-breaking perturbations, and remains effective even for time-dependent noise in the quasi-static limit~\cite{SupMat_arxiv}.

\paragraph{GHZ state preparation in Rydberg atom arrays. --}
As a second, experimentally relevant example, we turn to Rydberg atom arrays~\cite{Browaeys2020}, a paradigmatic quantum platform that naturally features coherent control and strong interactions with static imperfections~\cite{wurtz2023aquila}.
Each atom is modeled as a two-level system with ground state $\ket{g}$ and Rydberg state $\ket{r}$, and the dynamics is described by the Hamiltonian  
\begin{equation}
\hat{H}_{\mathrm{Ryd}} = \sum_j \left( \frac{\Omega }{2} \hat{\sigma}_j^x - \Delta \hat{n}_j \right) + \sum_{j<k} U_{jk} \hat{n}_j \hat{n}_k,
\label{eq:H_ryd}
\end{equation}
where  $\hat{\sigma}_j^x = \ket{g}_j \! \bra{r} + \ket{r}_j \! \bra{g}$ and $\hat{n}_j = \ket{r}_j \! \bra{r}$.  
Here, $\Omega$ is the Rabi frequency, $\Delta$ is the detuning, and $U_{jk} \sim 1/|\mathbf{x}_j - \mathbf{x}_k|^6$ is the van der Waals interaction between atoms at positions $\mathbf{x}_j$ and $\mathbf{x}_k$. The blockade radius $R_b$ is defined as the characteristic length scale at which $U_{R_b} = \Omega$, marking the distance below which simultaneous Rydberg excitations are suppressed \cite{Jaksch2000,Lukin2001,Gaëtan2009}.

We consider arrays of $16$ atoms on a square lattice~\cite{Samajdar2020,Ebadi_2021} with open boundary conditions, since this geometry is readily realizable experimentally and avoids the need for additional local control required in open chains~\cite{Omran2019}.
Similarly to the Ising model, $\hat{H}_{\rm Ryd}$ possesses a $\mathbb{Z}_2$ symmetry encoded as a lattice reflection exchanging the two sublattices, and, for suitable $R_b$, exhibits a disordered and an ordered phases, separated by a critical point. Below, we set $R_b = 1.15 a$, where $a$ is the lattice spacing. 

\begin{figure}[t]
    \phantomsubfloat{\label{fig3:a}}
    \phantomsubfloat{\label{fig3:b}}
    \centering  
    \includegraphics[width=1.0\columnwidth]{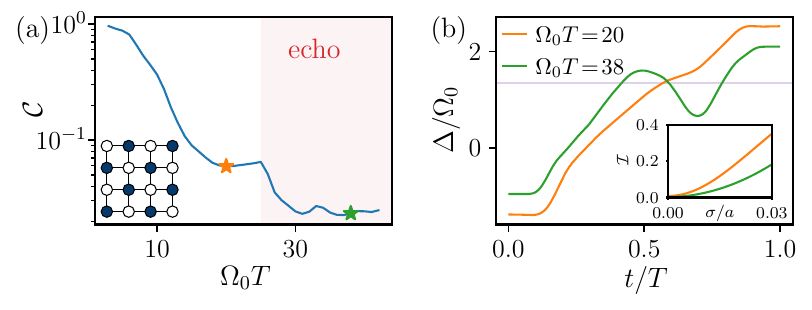}
    \vspace*{-5mm}
    \caption{ \textbf{GHZ state preparation in Rydberg arrays.} (a)~Optimal cost function \cref{eq:cost_f} vs total preparation time, for $N_{\rm s} = 30$ and $\sigma=0.01 a$. In the shaded red region the optimal protocols exhibit the features of the echo mechanism. (b)~Optimal profiles for $\Omega_0 T=20$ (standard adiabatic) and $\Omega_0 T = 38$ (adiabatic echo). The insets compare the robustness of the two against the strength of the positional disorder. The shaded purple region marks the finite-size critical regime, estimated from two independent ground state diagnostics~\cite{SupMat_arxiv}.
        }
    \label{fig3}
\end{figure}

We aim to prepare $\ket{\psi^+_{\rm GHZ}} = \frac{1}{\sqrt{2}} \left( \ket{g}_{\! A} \! \ket{r}_{ \! B} + \ket{r}_{\! A} \!  \ket{g}_{ \! B} \right)$, where $\ket{g}_{\! A} \!  \ket{r}_{ \! B}$ denotes the configuration with all atoms in sublattice $A$ ($B$) in state $\ket{g}$ ($\ket{r}$), and vice versa.  
We initialize the system in the product state $\ket{\psi_0} = \ket{g g \cdots g}$ and evolve it under the time-dependent Rydberg Hamiltonian in \cref{eq:H_ryd}, with $\Omega(t) = \Omega_0 w(t)$, where $w(t)$ is a window function satisfying $w(0) = w(T) = 0$~\cite{SupMat_arxiv}. We use the detuning $\Delta(t)$ as the control function. Provided that $\Delta(0) < 0$, $\Delta(T) > 0$, and that $\mathbb{Z}_2$ symmetry is not explicitly broken, $\ket{\psi_0}$ is the ground state for $t = 0$, while $\ket{\psi^+_{\rm GHZ}}$ is the ground state for $t \to T$.

A relevant source of symmetry breaking in this setting arises from static but spatially inhomogeneous atomic displacements, which vary from shot to shot and explicitly break the $\mathbb{Z}_2$ lattice reflection symmetry (cf.~\cref{fig1:d}).  
These experimental imperfections are particularly significant due to the diagonal interaction $U_{\sqrt{2}a} \! \simeq \! 0.3\, \Omega$, which is comparable to the Rabi frequency.

To identify the echo protocol for robust GHZ state preparation, we apply GRAPE to the cost function \cref{eq:cost_f} (cf.~\cref{fig3}), where each sample consists of one displacement $\delta \mathbf{x}_j$ per atom, independently drawn from a gaussian distribution with zero mean and standard deviation $\sigma = 0.01 a$.  
As in the Ising model, the optimal control solution differs qualitatively from a standard adiabatic protocol beyond a time scale $\Omega_0 T^* \simeq 20$ (colored region in \cref{fig3:a}).  
The corresponding optimal profiles are shown in \cref{fig3:b}: while for $T < T^*$ the solution features a monotonic ramp-up of $\Delta(t)$, for $T>T^*$ the optimal protocol exhibits the characteristic structure of the echo scheme, featuring three sweeps across the phase transition (cf.~\cref{fig3:b}).  
The inset of \cref{fig3:b} further highlights the enhanced robustness of the echo protocol, which was experimentally validated in a parallel study on Rydberg ladders~\cite{Senoo2025,SupMat_arxiv}, enabling the preparation of GHZ states nearly twice as large as those achievable with standard adiabatic protocols.

\begin{figure}[t]
    \phantomsubfloat{\label{fig4:a}}
    \phantomsubfloat{\label{fig4:b}}
    \centering  
    \includegraphics[width=1.0\columnwidth]{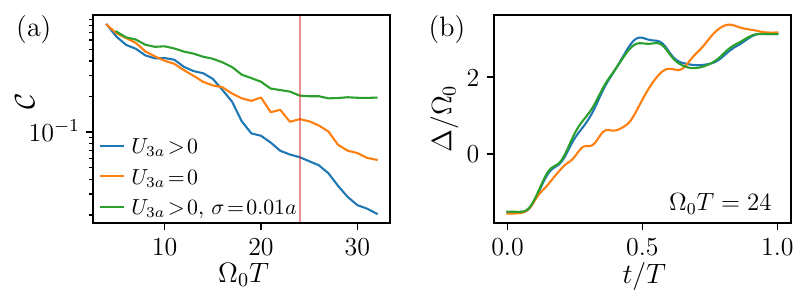}
    \vspace*{-5mm}
    \caption{ 
        \textbf{RVB state preparation in Rydberg arrays.}
        (a)~Optimal RVB preparation fidelity vs total preparation time on the ruby lattice with $24$ atoms, truncating ($U_{3 a} = 0$) and including ($U_{3a} > 0$) the van der Waals potential. For $U_{3a} > 0$ the optimization is also performed including positional disorder via the cost function \cref{eq:cost_f}, averaged over $N_{\rm s} = 30$ gaussian distributed samples (green line). (b)~Optimal control profiles for $\Omega_0 T = 24$ (vertical red line in (a)). The optimal protocol exhibits the structure of the adiabatic echo only when $U_{3a} > 0$.  
        }
    \label{fig4}
\end{figure}

\paragraph{Quantum spin liquid state preparation. --} 
As a final example, we consider Rydberg atoms on the ruby lattice and set $R_b = 2.3a$~\cite{Verresen2021,Semeghini2021}. Similarly to the GHZ preparation discussed above, we fix $\Omega(t) = \Omega_0 w(t)$, with $w(0) = w(T) = 0$~\cite{SupMat_arxiv}, take $\Delta(t)$ as the control function, with $\Delta(0) < 0$ and $\Delta(T) > 0$, and initialize the system in the product state $\ket{g g \cdots g}$. For $t = T$, and with the van der Waals potential truncated as $U_{R} = 0$ for $R > \sqrt{7}a$, the classical ground states correspond to all fully-packed dimer coverings of the kagome lattice~\cite{Verresen2021}. Their equal-weight superposition--the resonating valence bond (RVB) state~\cite{Anderson1973}--is a $\mathbb{Z}_2$ quantum spin liquid~\cite{Misguich2002,Moessner2011}, and is the target of our preparation. Previous works have demonstrated that this state can be prepared with high fidelity in this setting~\cite{Giudici2022,Sahay2022,Gjonbalaj2025}.
However, the inclusion of longer-range interaction tails $U_{3 a}$ lifts the degeneracy of the RVB components in the classical limit, imprinting relative phases during the evolution and thereby degrading the fidelity of the prepared state~\cite{Giudici2022,Zeng2025}.

While this perturbation does not explicitly break any microscopic symmetry of the Rydberg Hamiltonian on the ruby lattice, it plays a role analogous to the static perturbation $\hat{V}$ in the GHZ preparation discussed earlier, with the two GHZ components replaced by exponentially many RVB configurations. To determine the optimal protocol for preparing the RVB state in this setting, we apply GRAPE to the full Hamiltonian \cref{eq:H_ryd} on a periodic cluster of 24 atoms, including all van der Waals interaction tails. Remarkably, as shown in \cref{fig4:b}, the resulting optimal control profile displays the characteristic non-monotonic shape of the adiabatic echo. 
This solution is also robust, as the optimal protocol remains essentially unchanged when positional disorder is included in the optimization (cf. blue and green lines in \cref{fig4:b}).
In contrast, when the interaction is truncated at distance $R = \sqrt{7}a$, the optimal solution is monotonic (cf. orange line in \cref{fig4:b}), confirming that the echo protocol arises specifically to mitigate the effect of the long-range tails. 

\paragraph{Conclusions. --}
We introduced the adiabatic echo protocol as a general approach to many-body quantum state preparation in the presence of static perturbations. Its key feature is the suppression of leading-order errors via dynamically engineered interference, which we analyzed both analytically and numerically. Going beyond standard approaches to many-body optimal control, which rely on restrictive parametrizations of the control fields~\cite{Caneva2011,Cui2017,Omran2019,SupMat_arxiv}, we applied GRAPE to various interacting models and uncovered robust solutions that naturally exhibit the echo structure.

This work demonstrates that unbiased optimal control methods such as GRAPE can be effectively applied to strongly interacting quantum systems, highlighting their effectiveness in this complex setting. The adiabatic echo protocol introduced here offers a novel approach to robust quantum state preparation, with potential applicability across a wide range of platforms beyond those explicitly considered. Taken together, these findings open new directions for theoretical and numerical exploration of robust many-body quantum dynamics.

\let\oldaddcontentsline\addcontentsline
\renewcommand{\addcontentsline}[3]{}
\begin{acknowledgments}
\paragraph{Acknowledgments. --} We thank C. Fromonteil and P. Zoller for helpful discussions.
This work is supported by the ERC Starting grant QARA (Grant No. 101041435), the Horizon Europe programme HORIZON-CL4-2022-QUANTUM02-SGA via the project 101113690 (PASQuanS2.1) and by the Austrian Science Fund (FWF) (Grant No. DOI 10.55776/COE1). G. G. acknowledges support from the European Union’s Horizon Europe program under the Marie Sk\l{}odowska Curie Action TOPORYD (Grant No. 101106005). A.B. acknowledges support by the Swiss National Science Foundation under grant No. 222216. A.M.K. acknowledges support by ONR (N00014-23-1-2533), AFOSR (FA9550-23-1-0097), and NIST.
\end{acknowledgments}

\paragraph{Data availability. --} The data that support the findings of this article are openly available \cite{zeng_2026_18865750}.

\bibliography{library}	
\let\addcontentsline\oldaddcontentsline

\clearpage
\onecolumngrid
\begin{center}
    \textbf{\Large Supplementary Material}
\end{center}
\normalsize

\setcounter{equation}{0}
\setcounter{figure}{0}
\setcounter{table}{0}
\makeatletter
\renewcommand{\theequation}{S\arabic{equation}}
\renewcommand{\thefigure}{S\arabic{figure}}
\setlength\tabcolsep{10pt}
\setcounter{secnumdepth}{2}

\newcommand\numberthis{\addtocounter{equation}{1}\tag{\theequation}}
\newcommand{\insertimage}[1]{\includegraphics[valign=c,width=0.04\columnwidth]{#1}}

\tableofcontents

\section{Preparation infidelity from time-dependent perturbation theory}
\label{app:Perturbation}

In this section we derive the expression for the preparation infidelity used in the main text.\\
We consider the time-dependent Hamiltonian $\hat{H}(s(t)) = \hat{H}_0 (s(t)) + \epsilon \hat{V} $. We want to prepare the ground state of $\hat{H}_0(s(T))$, i.e., $\ket{\psi_{\mathrm{target}}} = \ket{E_0(s(T))}$, starting from the ground state $\ket{E_0(s(0))}$ of $\hat{H}_0(s(0))$. To lighten notation, we drop the explicit dependence on the control parameter $s$.
The preparation infidelity is defined as
\begin{equation}
    \mathcal{I} = 1 - \left| \braket{ \psi_{\mathrm{target}} | \psi(T) } \right|^2 = 1 - \left|  \braket{ E_0(T) | \hat{U}(T) | E_0(0) } \right|^2 = \sum_{n \neq 0 } \left|  \braket{ E_n(T) | \hat{U}(T) | E_0(0) } \right|^2 ,
    \label{eq:infidelity}
\end{equation}
where in the last step we used the completeness of the eigenbasis $\{\ket{E_n(T)}\}$.\\
In the interaction picture, the full time-evolution operator decomposes as $\hat{U}(T) = \hat{U}_0(T) \hat{U}_I(T)$, where:
\begin{equation}
    \hat{U}_I(T) = \mathrm{exp} \left( -i \epsilon \int_{0}^T d t \, \hat{U}^\dagger_0(t) \hat{V} \hat{U}^{\phantom{\dagger}}_0(t) \right) \simeq 1 - i \epsilon \int_{0}^T d t \, \hat{U}^\dagger_0(t) \hat{V} \hat{U}^{\phantom{\dagger}}_0(t) .
\end{equation}
Assuming adiabatic evolution, we approximate the unperturbed dynamics as
\begin{equation}
    \hat{U}_0(t) \ket{E_n(0)} \simeq \mathrm{exp} \left( -i \int_0^t d\tau  E_n(\tau) + i \int_0^t d \tau \braket{ E_n(\tau) | \partial_\tau E_n(\tau) } \right) \ket{E_n(t)} =: e^{ i \phi_n(t) } \ket{E_n(t)} , 
\end{equation}
such that \cref{eq:infidelity} becomes:
\begin{align}
       \mathcal{I} & = \sum_{n \neq 0 } \left|  \braket{ E_n(T) | \hat{U}_0(T) \hat{U}_I(T) | E_0(0) } \right|^2 \simeq \sum_{n \neq 0 } \left|  e^{i \phi_n (T) } \braket{ E_n(0) |  \hat{U}_I(T) | E_0(0) } \right|^2 \simeq \\ 
       &  \simeq \epsilon^2 \sum_{n \neq 0 } \left| \int_0^T d t   \braket{ E_n(0) |  \hat{U}^\dagger_0(t) \hat{V} \hat{U}^{\phantom{\dagger}}_0(t) | E_0(0) }  \right|^2 = \epsilon^2  \sum_{n \neq 0 } \left| \int_0^T d t   \braket{ E_n(t) |  \hat{V} | E_0(t) }  e^{-i(\phi_n(t) - \phi_0(t) ) } \right|^2  . 
\end{align}
We now observe that the infidelity $\mathcal{I}$ is invariant under local phase redefinitions of the instantaneous eigenstates $\{ \ket{E_n(t)} \}$. We exploit this freedom by choosing a gauge in which the Berry connection vanishes along the preparation path: $\braket{ E_n(\tau) | \partial_\tau E_n(\tau) }=0$. Using the identity $\int_0^t d\tau E_n(\tau) = \int_0^T d \tau E_n(\tau) -  \int_t^T d \tau E_n(\tau) $, and retaining only the dominant contribution from the first excited state, the preparation infidelity at leading order becomes:
\begin{equation}
    \mathcal{I} \simeq \epsilon^2 \left| \int_0^T d t   \braket{ E_1(t) |  \hat{V} | E_0(t) }  e^{-i \int_t^T d \tau \, ( E_1(\tau) - E_0(\tau) ) } \right|^2 .
    \label{eq:infidelity_SM}
\end{equation}

\section{Robustness of adiabatic protocols: symmetry-breaking vs non–symmetry-breaking perturbations}

In the main text, we showed that symmetry-breaking perturbations can have a strong detrimental impact on state-preparation fidelities, motivating the need for protocols that are robust against such effects. In this section, we demonstrate that symmetry-preserving perturbations, by contrast, have a much milder influence on the fidelity when preparing a symmetric many-body state such as the GHZ state. Furthermore, we show that the robustness of the adiabatic echo protocol against symmetry-breaking perturbations is not limited to the specific perturbation considered in the main text but extends more generally.

\begin{figure}[h!]
    \phantomsubfloat{\label{fig:App_Ising_different_perturbationa}}
    \phantomsubfloat{\label{fig:App_Ising_different_perturbationb}}
    \centering  
    \includegraphics[width=0.55\columnwidth]{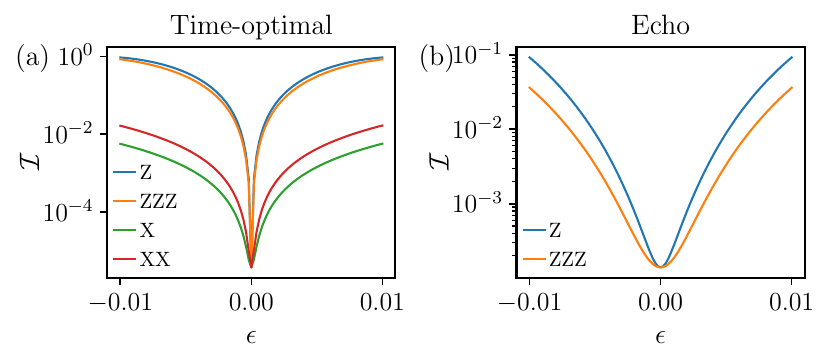}
    \vspace*{-3mm}
    \caption{GHZ state preparation infidelity $\mathcal{I}$ in the Ising chain under different perturbations, as a function of the perturbation strength $\epsilon$. (a)~The standard adiabatic time-optimal protocol (black line in Fig.~2b of the main text) is robust against symmetry-preserving perturbations ($\hat{V}^X$ and $\hat{V}^{XX}$), but sensitive to symmetry-breaking perturbations ($\hat{V}^Z$ and $\hat{V}^{ZZZ}$). 
(b)~The adiabatic echo protocol (colored lines in Fig.~2b of the main text) remains robust against both $\hat{V}^Z$ and $\hat{V}^{ZZZ}$.}
    \label{fig:App_Ising_different_perturbation}
\end{figure}

We consider the Ising chain in a transverse field subject to a generic perturbation:  
\begin{equation}
\hat{H}_{\mathrm{Ising}} = - J \sum_j \hat{\sigma}_j^z \hat{\sigma}_{j+1}^z - g \sum_j \hat{\sigma}_j^x + \epsilon \sum_j \hat{V}_j \, .
\label{eq:ising_pert}
\end{equation}
We analyze the effect of different perturbations $\hat{V}_j$ on the two GHZ state preparation protocols shown in Fig.~2b of the main text: the standard adiabatic time-optimal protocol (black line) and the adiabatic echo protocol (colored lines), both obtained from optimal control as described in the main text.  
In \cref{fig:App_Ising_different_perturbation}, we plot the GHZ infidelity $\mathcal{I}$ as a function of the perturbation strength $\epsilon$ for four different choices of $\hat{V}_j$: two that preserve the $\mathbb{Z}_2$ symmetry ($\hat{V}^X_j = \hat{\sigma}^x_j$ and $\hat{V}^{XX}_j = \hat{\sigma}^x_j \hat{\sigma}^x_{j+1}$) and two that break it ($\hat{V}^Z_j = \hat{\sigma}^z_j$ and $\hat{V}^{ZZZ}_j = \hat{\sigma}^z_{j-1} \hat{\sigma}^z_j \hat{\sigma}^z_{j+1}$). 
\cref{fig:App_Ising_different_perturbationa} shows that the $\mathcal{I}$ obtained from the time-optimal protocol is already resilient against perturbations that preserve the symmetry, while \cref{fig:App_Ising_different_perturbationb} demonstrates that the echo protocol maintains its performance under both symmetry-breaking perturbations. 
Remarkably, the resilience against $\hat{V}^{ZZZ}$ requires no additional optimization of the control profile, which was obtained by including only $\hat{V}^Z$ in the cost function of Eq.~(4) in the main text. 
This demonstrates the generality and intrinsic error-cancellation mechanism of the echo protocol.

\section{Contribution to the preparation infidelity from disordered and ordered phases}

\begin{figure}
    \phantomsubfloat{\label{fig:App_Va}}
    \phantomsubfloat{\label{fig:App_Vb}}
    \phantomsubfloat{\label{fig:App_Vc}}
    \centering  
    \includegraphics[width=0.8\columnwidth]{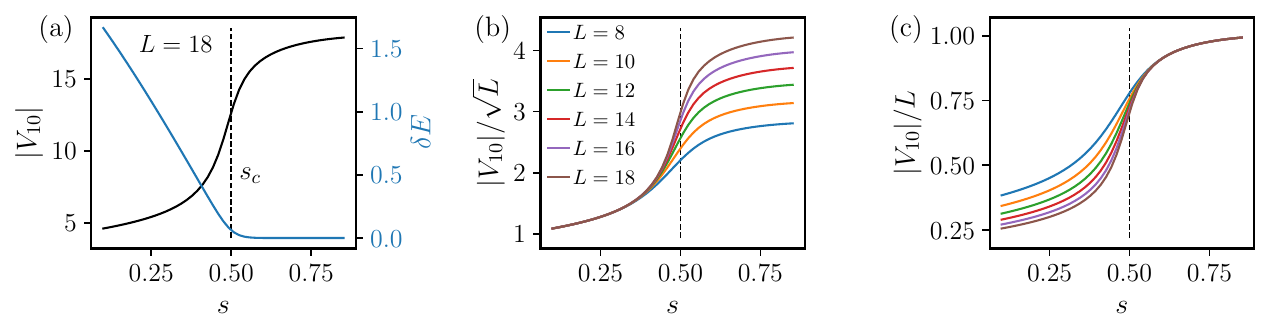}
    \vspace*{-3mm}
    \caption{(a)~Symmetry-breaking matrix element $|V_{10}|$ (black) and lowest energy gap $\delta E$ (blue) as functions of the control parameter $s$ (cf. Eq.~(3) of the main text and discussion below) for $L=18$. The vertical line marks the critical point $s_c = 1/2$. (b,c)~Rescaling $|V_{10}|$ by $\sqrt{L}$ (b) and $L$ (c) results in curve collapse for $s<s_c$ and $s>s_c$, respectively, demonstrating $V_{10}(s<s_c) \sim \sqrt{L}$ and $V_{10}(s>s_c) \sim L$.   }
    \label{fig:App_V}
\end{figure}

\begin{figure}
    \phantomsubfloat{\label{fig:App_V_Rydberga}}
    \phantomsubfloat{\label{fig:App_V_Rydbergb}}
    \phantomsubfloat{\label{fig:App_V_Rydbergc}}
    \centering  
    \includegraphics[width=0.8\columnwidth]{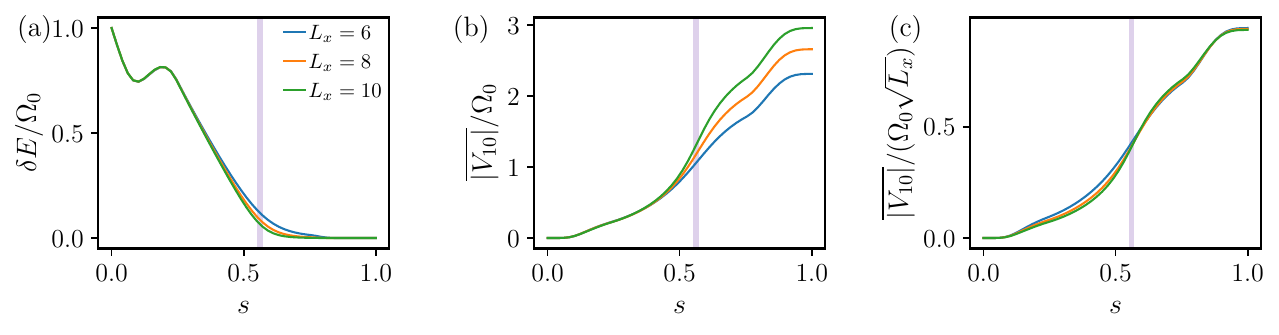}
    \vspace*{-3mm}
    \caption{(a)~Lowest energy gap $\delta E / \Omega_0$ as a function of the control parameter $s$ for the Rydberg Hamiltonian on a ladder (cf. Eq~(5) of the main text) with $\Omega(s) = \Omega_0 \, w(s)$ (cf. \cref{fig:App_window}) and $\Delta(s) = \Omega_0 ( -1 + 4s)$, for various ladder lengths $L_x$. The vertical shaded region marks the phase transition regime discussed in \cref{sec:phase_transition}. (b,c)~Curve collapse of the symmetry-breaking matrix element $\overline{V_{10}}(s)/ \Omega_0$, averaged over $1000$ realizations of the positional disorder. For $s<s_c$, $\overline{V_{10}}(s)/ \Omega_0$ is independent of $L_x$ (b), while for $s>s_c$ the data collapse shows $\overline{V_{10}}(s)/ \Omega_0 \sim \sqrt{L_x}$ (c). }
    \label{fig:App_V_Rydberg}
\end{figure}

In the main text, we neglected the contribution to the preparation infidelity coming from regions where the control field $s(t)$ lies in the disordered phase of the Hamiltonian $\hat{H}(s)$, i.e., $s < s_c$.
To justify this approximation, we divide the protocol into segments separated by crossings of the critical point $s_c$, so that within each segment, either $s(t) < s_c$ or $s(t) > s_c$ holds throughout. We label the segments by $k$, and approximate $V_{10}(s) \simeq V_{10}^k$ and $\delta E(s) \simeq \delta E_k$ within each one.
From Eq.~(1) of the main text, the infidelity becomes
\begin{equation}
\mathcal{I} \simeq \epsilon^2 \left| \sum_k e^{i \alpha_k} \, V_{10}^k \, \frac{e^{i \delta E_k T_k} - 1}{\delta E_k} \right|^2 = \epsilon^2 \left| \sum_k A_k \right|^2 ,
\end{equation}
where $T_k$ is the time spent in interval $k$ and $\alpha_k$ is the dynamical phase accumulated before interval $k$. 

In the trivial phase ($s<s_c$), $\delta E_k$ is finite, so $A_k \sim e^{i \alpha_k} V^k_{10} / \delta E_k$. In the ordered phase ($s>s_c$), by contrast, the gap $\delta E_k$ is exponentially small in the system size $L$, such that $\delta E_k T_k \ll 1$, yielding $A_k \simeq e^{i \alpha_k} V^k_{10} \, T_k$.
Since adiabaticity requires $T_k \sim L$, we have $A_k (s>s_c) \sim V^k_{10} \, L $, while $A_k (s<s_c) \sim V^k_{10}$. Moreover, as we show below in two explicit examples, the matrix element $V_{10}(s < s_c)$ is parametrically suppressed with system size compared to $V_{10}(s > s_c)$.

In the Ising model, $\hat{V} = \sum_j \hat{\sigma}^z_j$, and the system-size scaling of $V_{10}$ can be computed analytically for both $s=0$ and $s=1$. For $s=0$, corresponding to the disordered phase, the ground state is $\ket{E_0} = \ket{ + + \dots +}$ and the first excited state is $\ket{E_1} = \frac{1}{\sqrt{L}}\sum_j \hat{\sigma}^z_j \ket{E_0}$, yielding $\braket{ E_1 | \hat{V} | E_0 } = \sqrt{L}$. For $s=1$, deep in the ordered phase, the ground and first excited states are $\ket{E_0} = \frac{1}{\sqrt{2}} ( \ket{ \uparrow \uparrow \dots \uparrow} + \ket{\downarrow \downarrow \dots \downarrow} )$ and $\ket{E_1} = \frac{1}{\sqrt{2}} ( \ket{ \uparrow \uparrow \dots \uparrow} - \ket{\downarrow \downarrow \dots \downarrow} )$, respectively, leading to $\braket{ E_1 | \hat{V} | E_0 } = L$. As shown in \cref{fig:App_V}, these distinct scaling behaviors persist throughout their respective phases, leading to additional system-size suppression of the amplitude $A_k$ in the disordered regime: $A_k(s<s_c)/A_k(s>s_c) \sim 1/L^{3/2}$.

We observe the same scaling behavior in Rydberg atom arrays. In this case, the symmetry-breaking perturbation takes the form
\begin{equation}
\hat{V} = \frac{a}{\sigma} \, \Omega_0 \sum_{j<k} \left(  \frac{ R_b^6 }{ \left| ( \mathbf{x}_j + \delta \mathbf{x}_j ) - ( \mathbf{x}_k + \delta \mathbf{x}_k ) \right|^6 } -  \frac{ R_b^6 }{ \left|  \mathbf{x}_j - \mathbf{x}_k \right|^6 } \right),
\end{equation}
where $\delta \mathbf{x}_j$ is a gaussian-distributed displacement with zero mean and standard deviation $\sigma$, and the prefactor is chosen so that $\hat{V}$ is independent of $\sigma/a$ for $\sigma/a \ll 1$.
We focus on the ladder geometry and evaluate $|V_{10}(s)|$ for the Rydberg Hamiltonian (Eq.~(5) of the main text), averaging over 1000 disorder realizations. Here $0 \leq s \leq 1$, with $\Omega(s) = \Omega_0 \, w(s)$ (see \cref{fig:App_window}) and $\Delta(s) = \Omega_0 (-1 + 4s)$. The resulting $\overline{V_{10}}(s)$ is shown in \cref{fig:App_V_Rydberg}, where the overline denotes the disorder average. We find $\overline{V_{10}}(s)/ \Omega_0 \sim \mathrm{const}$ for $s < s_c$, while $\overline{V_{10}}(s)/ \Omega_0 \sim \sqrt{L_x}$ for $s > s_c$, where $L_x$ is the linear size of the ladder. As in the Ising case, this implies $A_k(s < s_c)/A_k(s > s_c) \sim 1/L_x^{3/2}$.

\section{Optimal control via GRAPE}\label{app:OptimalControl}

In this section, we review the optimal control methods used to optimize the profile of the control parameter $s(t)$. The cost function is defined as the infidelity between the prepared state and the target state:
\begin{equation}\mathcal{C} = 1-|\langle\psi_{\rm target}|\hat{U}(T)|\psi_0\rangle|^{2} \simeq 1-|\langle\psi_{\rm target}|\prod_{j=0}^{N-1}\hat{U}_j \,|\psi_{0}\rangle|^{2},
\end{equation}
where the time-evolution operator is discretized into $N$ steps with $dt = T/N$, such that  $\hat{U}_j =\exp\left(-i\hat{H}(s_j)dt\right)$, $s_j=s(t_j)$, and $t_j = (j+\frac{1}{2}) dt$, where $j=0,\dots N-1$. In this work, we used $N=150$.\\
We employ a gradient descent algorithm to minimize $\mathcal{C}$. The gradient of the cost function with respect to the control parameter $s_{j}$ at each time is 
\begin{equation}
\frac{\partial \mathcal{C}}{\partial s_{j}} =2\mathrm{Re} \left( \langle\psi(T)|\psi_{\rm target}\rangle \langle\psi_{\rm target}| \Bigg( \prod_{k=j+1}^{N-1} \! \! \hat{U}_k \Bigg) \frac{d \hat{U}_j }{d s_{j}}  \Bigg( \prod_{k=0}^{j-1} \hat{U}_k \Bigg) |\psi_0 \rangle\right).
\label{eq:grad_i}
\end{equation}
The derivative $\frac{d \hat{U}_j }{d s_j}$ has to be treated with care when $\hat{H}(s_j)$ and $\frac{d \hat{H} (s_j) }{d s_j}$ do not commute. One has
\begin{equation}
    \frac{d \hat{U}_j }{d s_j} = \underset{ h \to 0 }{ \lim } \frac{ \exp \left( -i d t \hat{H}(s_j + h ) \right) - \exp \left( -i d t \hat{H}(s_j) \right) }{h} = \underset{ h \to 0 }{ \lim } \frac{ \exp \left( -i d t \left( \hat{H}(s_j) + h \frac{d \hat{H} (s_j) }{d s_j} \right) \right) - \exp \left( -i d t \hat{H}(s_j) \right) }{h}.
\end{equation}
For an exact numerical evaluation of the r.h.s. one can use the formula~\cite{AlMohyHigham2009}
\begin{equation}
    f\left(
\begin{bmatrix}X & E\\
0 & X
    \end{bmatrix} \right) = \begin{bmatrix} f(X) & Df(X,E) \\
0 & f(X)
    \end{bmatrix},
    \qquad 
    \mathrm{where} \quad 
    Df(X,E) = \lim_{h\rightarrow0}\frac{f(X+hE)-f(X)}{h},
    \label{eq:frechet}
\end{equation}
with $f = \exp$, $X = -i dt \hat{H}(s_j)$, and $E = -i dt \frac{d \hat{H}(s_j)}{d s_j}$. \cref{eq:frechet} is amenable to efficient computation via Krylov-based algorithms for sparse matrices or tensor network methods based on the time-dependent variational principle (TDVP)~\cite{Haegeman2016}. In practice, for the time-steps and system sizes considered in this work, we employed the simple approximation
\begin{equation}
    \frac{d \hat{U}_j}{d s_j}\simeq -idt\frac{d \hat{H}(s_j)}{d s_{j}}\hat{U}_j,
    \label{eq:approx_gradient}
\end{equation}
which can be straightforwardly computed with sparse matrix routines or any time-dependent tensor network method (see \cref{app:FSscaling}). \\
For the gradient-descent minimization, we used the L-BFGS-B algorithm from the SciPy package~\cite{scipy}. The optimization is performed iteratively, beginning with a short total evolution time $T$ and a linear profile as the initial guess. Upon convergence, we increase $T$ as $T=T+dT$ and use the previously optimized profile as the initial guess. This iterative strategy provides an efficient way to explore the dependence of the optimal cost function on the total preparation time $T$. At short $T$, the optimization landscape is simple and convergence is fast, while at larger $T$ it becomes increasingly complex and sensitive to the initial conditions. To accurately locate the crossover time where the echo solution first becomes advantageous, we also perform a reverse iterative procedure, starting from the optimal large-$T$ profile and gradually decreasing $T$.

In practical implementations, the control function $s(t)$ must vary without large jumps between consecutive time-steps to ensure experimental feasibility. To enforce this, we add a regularization term to the cost function that penalizes rapid variations of $s(t)$~\cite{Khaneja2005}:
\begin{equation}
    \mathcal{C}_\eta = \mathcal{C} + \eta \int_0^{1} du \left(\frac{ds}{du}\right)^2  \simeq \mathcal{C} + \eta N \sum_j \left(s(t_{j+1})-s({t_{j}}))\right)^2,
\end{equation}
where $u=t/T$ is the rescaled time and $\eta$ is the strength of the regularization. 
Throughout this work, we fix $\eta=10^{-3}$. \\
Additionally, one can impose boundary conditions $s(0) = s_0$ and $s(T) = s_f$ by adding a penalty term that fixes the initial and final values:
\begin{equation}
    \mathcal{C}_{\eta,\mu}=\mathcal{C}_\eta + \mu\left( (s(0) -s_{0})^2 + (s(T) - s_{f})^2\right),
\end{equation} 
for some constant $\mu \gg \mathcal{C}_0$.
In this work, we enforce boundary conditions $s(0)=0$ and $s(T)=1$ for the GHZ state preparation in the ferromagnetic quantum Ising chain, by setting $\mu=1$.

The discussion above extends by linearity to the evaluation of the gradient of the cost function in Eq.~(4) of the main text.

\begin{figure}[t]
    \centering  
    \includegraphics[width=0.25\columnwidth]{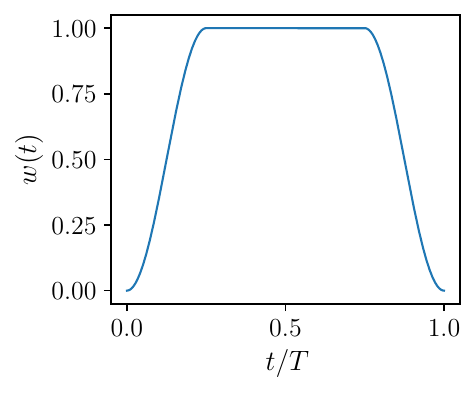}
    \vspace*{-3mm}
    \caption{Cosine-tapered window function \cref{eq:window} with $\alpha=0.25$.}
    \label{fig:App_window}
\end{figure}

For the Rydberg array preparation of GHZ and RVB states (cf. Fig.~3 and Fig.~4 of the main text), we take the detuning $\Delta(t)$ as the control function and parameterize the Rabi frequency as $\Omega(t) = \Omega_0 w(t)$, where $w(t)$ is a cosine-tapered window of total duration $T$ (cf. \cref{fig:App_window}):
\begin{equation}
    w(t) = 
    \begin{cases}
    \frac{1}{2} \left[ 1 + \cos \left( \frac{\pi t}{\alpha T} \right) \right], & 0\le t < \alpha T\\[1mm]
    1, & \alpha T \leq t \leq  \frac{T}{2} \\[1mm]
    w(T-t). & \frac{T}{2} < t
    \end{cases}
    \label{eq:window}
\end{equation}
Here, the parameter $\alpha$ sets the rise (and fall) time relative to the total preparation time. We use $\alpha = 0.25$ for the GHZ preparation and $\alpha = 0.15$ for the RVB preparation, chosen to minimize the infidelity in the regime $\Omega_0 T \lesssim 35$.\\
In principle, the Rabi frequency $\Omega(t)$ could also be treated as an independent control function. However, we find that doing so significantly slows convergence due to a more complex optimization landscape, while yielding only marginal improvements in fidelity.

\section{Echo protocol from alternative optimal control methods}\label{app:AlternativeOC}

In this section, we demonstrate that the echo protocol can also be obtained using alternative optimal control methods. 
Specifically, we employ a gradient-based approach that assumes a parametrized form for the control function, known as Gradient Optimization of Analytic conTrols (GOAT)~\cite{Machnes2018}. 
We consider the quantum Ising chain in a transverse field with a longitudinal field as the symmetry-breaking perturbation (see Eq.~(3) of the main text), and parametrize the control function as
\begin{equation}
    s(t) = \frac{t}{T} + \sum_{n=1}^{N_c} c_n \sin \left( \frac{ 2 \pi n \, t}{T}  \right),
    \label{eq:parametrization}
\end{equation}
where the coefficients $c_n$ ($n = 1, \dots, N_c$) are the variational parameters. 
This parametrization enforces the boundary conditions $s(0) = 0$ and $s(T) = 1$, as well as the symmetry $s(t) = s(T - t)$. The number of Fourier components $N_c$ determines the expressibility of the ansatz. 

The optimal control strategy closely follows the one outlined in the main text. We search for GHZ state preparation protocols $s(t)$ that are robust against symmetry-breaking perturbations by minimizing the cost function given in Eq.~(4) of the main text. The only difference with respect to the GRAPE algorithm of \cref{app:OptimalControl} lies in the computation of the gradient, which now reads:
\begin{equation}
\frac{\partial C}{\partial c_n}
= \sum_{j=0}^{N-1} \frac{\partial s_j}{\partial c_n} \frac{\partial C}{\partial s_j}
= \sum_{j=0}^{N-1} \sin \! \left(\frac{2 \pi n \, t_j}{T}\right) \frac{\partial C}{\partial s_j},
\end{equation}
where $\partial C / \partial s_j$ is given in \cref{eq:grad_i}, and $\partial s / \partial c_n$ follows from \cref{eq:parametrization}. The optimization is carried out using the L-BFGS-B algorithm from SciPy. 

\begin{figure}[h!]
    \phantomsubfloat{\label{fig:App_GOATa}}
    \phantomsubfloat{\label{fig:App_GOATb}}
    \phantomsubfloat{\label{fig:App_GOATc}}
    \phantomsubfloat{\label{fig:App_GOATd}}
    \centering  
    \includegraphics[width=0.99\columnwidth]{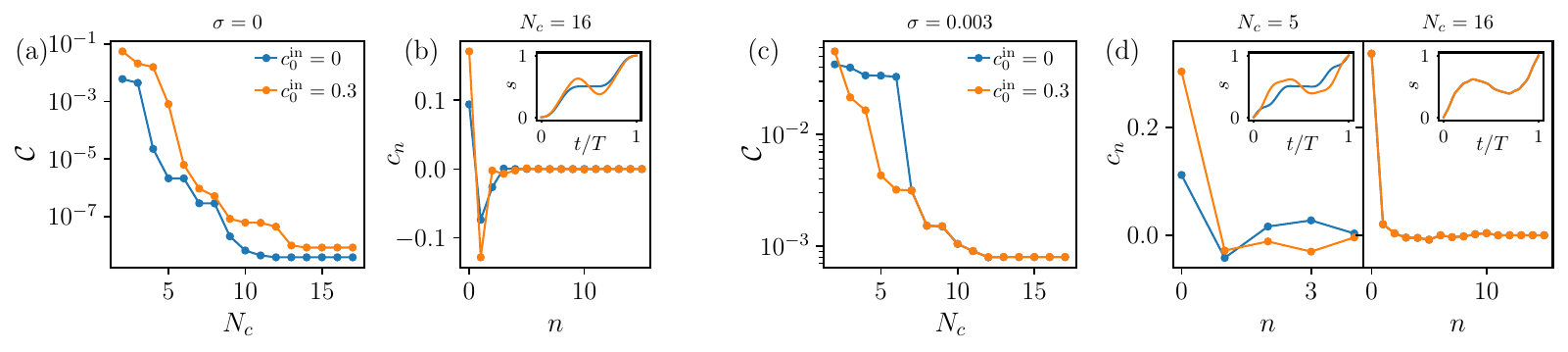}
    \vspace*{-3mm}
    \caption{Result of the GOAT optimization for GHZ state preparation in the Ising chain with $L = 16$, total time $T = 12$, and control function parametrization given in~\cref{eq:parametrization}. Panels~(a) and~(c) show the optimal cost function (Eq.~(4) of the main text) as a function of the number of Fourier components $N_c$ for $\sigma = 0$ and $\sigma = 0.003$, respectively. Two initial conditions are considered for $N_c = 1$: a monotonic $s(t)$ with $c_1^{\mathrm{in}} = 0$ (blue) and an echo-like $s(t)$ with $c_1^{\mathrm{in}} = 0.3$ (orange). For $\sigma = 0$ (no symmetry breaking), the two initial conditions converge to distinct solutions, as shown in panel~(b). For $\sigma = 0.003$, by contrast, both initial conditions ultimately converge to an echo-like optimal control profile, as shown in panel~(d), where the initially monotonic control becomes non-monotonic for $N_c > 5$. }
    \label{fig:App_GOAT}
\end{figure}

For a fixed total time $T = 0.8L$, we run the algorithm for a given $N_c$ and use the resulting coefficient vector $(c_1, c_2, \dots, c_{N_c})$ as the initial condition for the subsequent optimization with $N_c + 1$, setting $c_{N_c + 1} = 0$. Starting from $N_c = 1$, we iterate this procedure until the preparation infidelity reaches a plateau (cf.~\cref{fig:App_GOATa,fig:App_GOATc}). We consider two distinct initial conditions for $N_c = 1$, namely $c_1^{\mathrm{in}} = 0$ (linear ramp) and $c_1^{\mathrm{in}} = 0.3$ (echo-like initial condition), and two different values of the maximal symmetry-breaking strength $\sigma$ (cf.~main text below Eq.~(4)), $\sigma = 0$ (no symmetry breaking) and $\sigma = 0.003$. The outcome of this procedure is shown in \cref{fig:App_GOAT}.  

For $\sigma = 0$ (\cref{fig:App_GOATa,fig:App_GOATb}), the two initial conditions converge to distinct solutions as $N_c$ increases, both yielding comparable performance: a standard, monotonic adiabatic protocol for $c_1^{\mathrm{in}} = 0$ and an echo-like control protocol for $c_1^{\mathrm{in}} = 0.3$. This indicates that, at this time scale ($T/L = 0.8$), the echo protocol can compete with standard adiabatic protocols even in the absence of symmetry breaking. For $\sigma = 0.003$ (\cref{fig:App_GOATc,fig:App_GOATd}), by contrast, the optimization eventually converges to an echo protocol even when starting from a monotonic initial condition.

\section{Finite-size scaling of the GHZ state preparation in the quantum Ising chain}\label{app:FSscaling}

\begin{figure}[h!]
    \phantomsubfloat{\label{fig:App_Ising_FSSa}}
    \phantomsubfloat{\label{fig:App_Ising_FSSb}}
    \phantomsubfloat{\label{fig:App_Ising_FSSc}}
    \phantomsubfloat{\label{fig:App_Ising_FSSd}}
    \centering  
    \includegraphics[width=0.99\columnwidth]{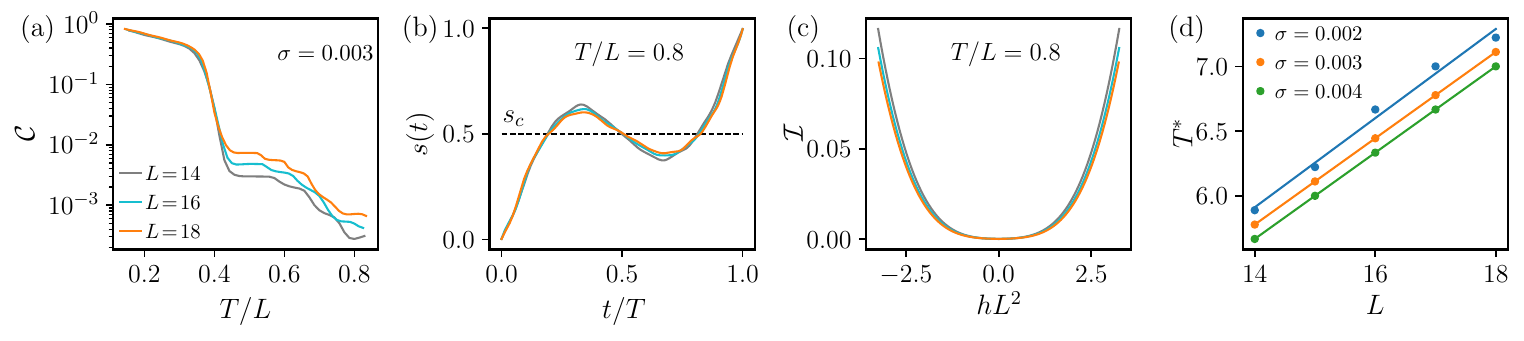}
    \vspace*{-3mm}
    \caption{(a)~Optimal cost function vs total preparation time $T$ for various system sizes $L$. The optimization is performed via the cost function Eq.~(4) of the main text, averaged over $N_{\rm s} = 30$ samples. (b)~Optimized control profiles $s(t)$ for fixed $T/L = 0.8$ while varying $L$.(c)~Corresponding preparation infidelity for the control protocols in (b), plotted against the scaled longitudinal field strength $h L^2$. (d) Finite-size scaling of the preparation time $T^*$ at which the echo protocol becomes advantageous, shown for different values of the maximal symmetry-breaking perturbation strength $\sigma$.}
    \label{fig:App_Ising_FSS}
\end{figure}

In the main text, we applied GRAPE to the cost function in Eq.~(4) to prepare the GHZ state in a quantum Ising chain of length $L = 18$, subject to a symmetry-breaking longitudinal field $h$. Here, we extend those results by performing a finite-size scaling analysis for system sizes $L = 14, 16, 18$, and by further scaling up the optimization of the echo protocol profile up to $L = 42$ using GRAPE in combination with tensor network methods.
\cref{fig:App_Ising_FSSa} shows the optimized cost function $\mathcal{C}$ (Eq.~(4) in the main text), averaged over $N_{\mathrm{s}} = 30$ perturbation samples, as a function of the total preparation time $T$. We observe a collapse of the curves when plotted against $T/L$, a hallmark of adiabatic protocols. In \cref{fig:App_Ising_FSSb}, we show that the optimal control profile $s(t)$ is essentially independent of $L$ when plotted at fixed $T/L$. 

In \cref{fig:App_Ising_FSSc} we plot the preparation infidelity $\mathcal{I}$ as a function of the rescaled perturbation strength $h L^2$. The approximate data collapse can be understood from the first-order expression of the infidelity in \cref{eq:infidelity_SM}. As discussed in the main text, the dominant contribution to $\mathcal{I}$ arises from the ordered phase, where $V_{10} \sim L$. Since $T \sim L$, one obtains $\mathcal{I}(h,L) \simeq h^2 T^2 (V_{10})^2 \sim h^2 L^4 = (h L^2)^2$. The validity of this scaling confirms that the echo protocol does not exactly cancel the first-order error due to the perturbation, while significantly suppressing its prefactor (cf.~Fig.~2c of the main text).

In \cref{fig:App_Ising_FSSd} we show the finite-size scaling of the preparation time $T^*$ at which the echo protocol becomes advantageous over a standard adiabatic protocol, for different values of the maximal symmetry-breaking strength $\sigma$ (cf.~Fig.~2 of the main text). The preparation time $T^*$ displays a clear linear dependence on the system size $L$, consistent with the adiabatic nature of the time evolution.

\begin{figure}[h!]
    \phantomsubfloat{\label{fig:App_Ising_TNa}}
    \phantomsubfloat{\label{fig:App_Ising_TNb}}
    \centering  
    \includegraphics[width=0.6\columnwidth]{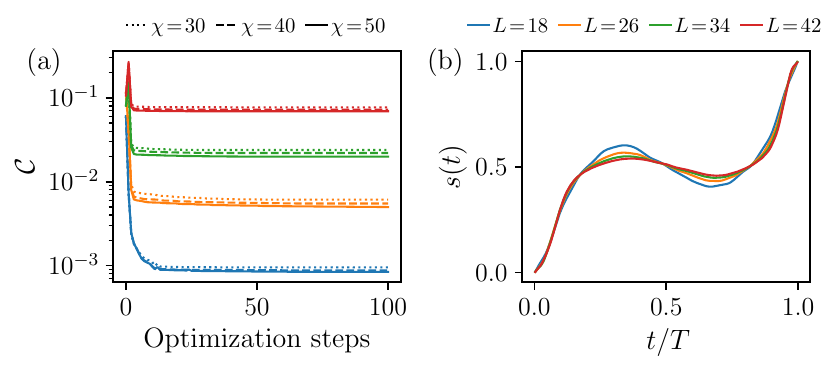}
    \vspace*{-3mm}
    \caption{(a) Cost function during the GRAPE optimization of the echo protocol in the quantum Ising chain for $T/L = 0.8$, shown for different system sizes $L$ and bond dimensions $\chi$. The approximate gradient is evaluated using \cref{eq:approx_gradient} within the TDVP tensor-network framework. (b) Finite-size scaling of the optimal control profiles for various system sizes.}
    \label{fig:App_Ising_TN}
\end{figure}

We can further scale up the optimization of the echo protocol up to $L=42$. We fix $T/L=0.8$ and run GRAPE by evaluating \cref{eq:approx_gradient} with TDVP. We iteratively increase the system size and use the converged solution for size $L$ as the initial condition for size $L+4$. The cost function during the optimization is shown in \cref{fig:App_Ising_TNa} for different bond dimensions $\chi$ and various system sizes $L$, demonstrating fast convergence during the optimization and with $\chi$. In \cref{fig:App_Ising_TNb}, we plot the converged echo protocol profile for different $L$. We note that the oscillation amplitude decreases with system size, consistent with the analytical argument presented in the main text: the integration interval increases with $L$, and the integral of the gap in region III must equal $\pi$ for the interference mechanism to occur.
These results confirm that the echo protocol remains stable upon scaling to larger systems, with its characteristic structure preserved and the optimization converging reliably across sizes. This demonstrates that the interference-based mechanism underlying the adiabatic echo is robust and persists for large $L$.

\section{Adiabatic echo protocol with GHZ state preparation on Rydberg ladders}\label{sec:Rydberg_ladder}

\begin{figure}[h!]
    \phantomsubfloat{\label{fig:App_ladder_a}}
    \phantomsubfloat{\label{fig:App_ladder_b}}
    \phantomsubfloat{\label{fig:App_ladder_c}}
    \centering
    \includegraphics[width=0.8\columnwidth]{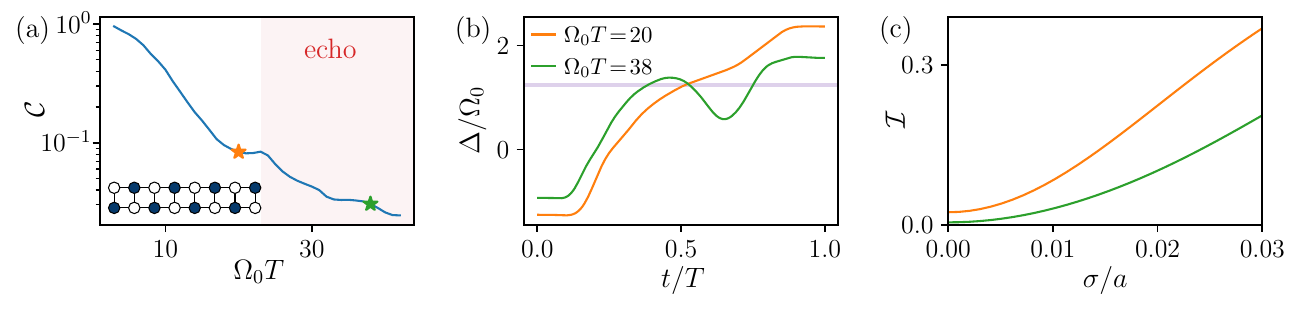}
    \vspace*{-3mm}
    \caption{GHZ state preparation on a Rydberg ladder with 16 atoms. (a)~Optimal cost function (cf. Eq.~(4)) as a function of  total preparation time, for $N_{\rm s} = 30$ samples, each consisting of one displacement $\delta \mathbf{x}_j$ per atom, gaussian distributed with standard deviation $\sigma=0.01 a$. In the shaded red region the optimal protocols acquire a profile typical of the echo mechanism. 
    (b)~Optimal control profiles for $\Omega_0 T=20$ (standard adiabatic) and $\Omega_0 T = 38$ (adiabatic echo). 
    (c)~Comparison of the robustness of the two protocols as a function of the standard deviation of the positional disorder.}
    \label{fig:App_Rydberg_ladder}
\end{figure}

In this section, we present results on GHZ state preparation in Rydberg atom arrays arranged in a two-leg ladder geometry~\cite{Eck2023,SotoGarcia2025,Zhang2025} (cf.~inset of \cref{fig:App_ladder_a}), which is the geometry used in the companion experimental work~\cite{Senoo2025}. As in the square lattice discussed in the main text, the ladder geometry exhibits a $\mathbb{Z}_2$ lattice reflection symmetry under exchange of the two sublattices, which is explicitly broken by shot-to-shot spatially inhomogeneous atomic displacements. 

We apply the same GRAPE optimization procedure described in the main text to identify robust GHZ state preparation protocols in the presence of positional disorder. The optimization results are shown in \cref{fig:App_Rydberg_ladder}. We consider open boundary conditions with blockade radius $R_b=1.15a$ and disorder strength $\sigma=0.01a$, matching the parameters used for the square lattice.

As shown in \cref{fig:App_ladder_a}, the optimal protocol exhibits the same qualitative behavior as in the square-lattice case: beyond the characteristic time scale $\Omega_0 T^* \simeq 20$, the optimal solution transitions from a standard adiabatic ramp to an adiabatic echo protocol. The corresponding optimal profiles and their robustness are shown in \cref{fig:App_ladder_b} and \cref{fig:App_ladder_c}, respectively, where the echo protocol shows enhanced robustness compared to the standard adiabatic protocol.

The above numerical findings have been used for the GHZ state preparation in an Ytterbium-171 tweezer array~\cite{Senoo2025}. Using the adiabatic echo protocol, we generate GHZ states exceeding the $50\%$ fidelity threshold on up to 20 atoms. In contrast, the maximum GHZ state size we could achieve with conventional adiabatic sweeps is 12 atoms. These results demonstrate the effectiveness of the echo protocol under realistic experimental conditions.

Furthermore, the echo protocol can be efficiently combined with other experimental techniques to further improve performance. In the same experiment, decay detection is enabled by a coherent state-transfer scheme that maps the prepared GHZ state from the interacting Rydberg manifold to a long-lived metastable qubit manifold. This suppresses decay errors and extends the accessible operation time of the system. As longer evolution times become available, static perturbations play an increasingly important role, further motivating the use of the echo protocol.

\section{Phase transition location in Rydberg atom arrays}
\label{sec:phase_transition}

\begin{figure}[h]
    \phantomsubfloat{\label{fig:App_Rydberg_transition_point_a}}
    \phantomsubfloat{\label{fig4:App_Rydberg_transition_point_b}}
    \centering  
    \includegraphics[width=0.64\columnwidth]{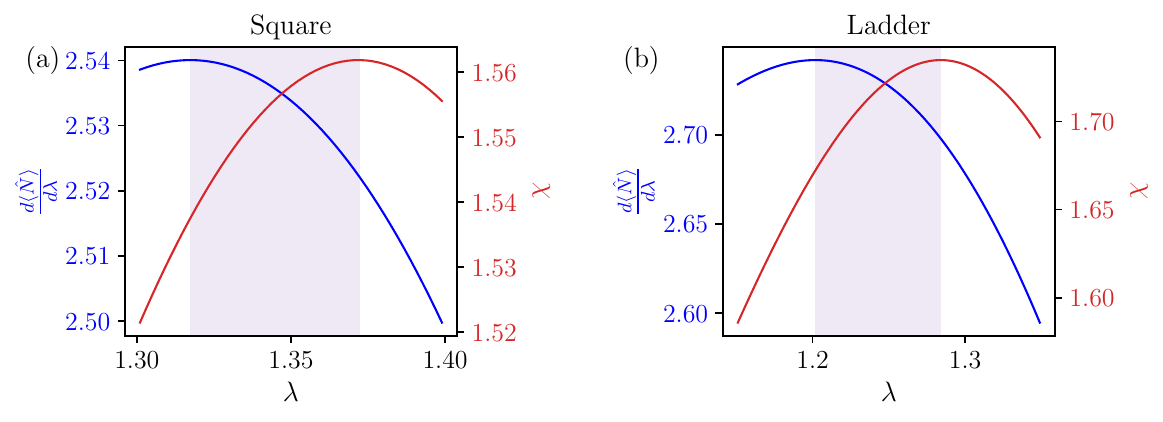}
    \vspace*{-3mm}
    \caption{ 
        Finite-size critical regime for arrays of $16$ Rydberg atoms with $R_b = 1.15 a$ on (a) a square lattice ($4 \times 4$) and (b) a ladder ($8 \times 2$). The shaded purple region marks the finite-size critical regime, and it is bounded by the peak of the excitation-number derivative (blue) and the peak of the fidelity susceptibility (red). 
        }
    \label{fig:App_Rydberg_transition_point}
\end{figure}

In the main text and \cref{sec:Rydberg_ladder}, we made use of the finite-size location of the transition point for Rydberg atoms on the square lattice and on the ladder, respectively. In Fig.~3b and \cref{fig:App_ladder_b}, we drew the critical regime as a purple horizontal narrow region, which we define as the interval between the transition points identified by two ground state diagnostics: (i) the derivative of the excitation number
\begin{equation}
    \frac{d}{d\lambda}\braket{ E_0(\lambda)|\hat{N}| E_0(\lambda) } \simeq \frac{\braket{ E_0(\lambda + \delta \lambda)|\hat{N}| E_0(\lambda + \delta \lambda) } - \braket{ E_0(\lambda)|\hat{N}| E_0(\lambda) } }{\delta \lambda},
\end{equation}
and (ii) the ground-state fidelity susceptibility \cite{Zanardi2006,Gu2010}
\begin{equation}
    \chi = -\frac{2}{\delta\lambda^2}\ln\left|\langle{ E_0(\lambda)}\ket{ E_0(\lambda + \delta\lambda)}\right|,
\end{equation}
where $\lambda=\Delta/\Omega$, $| E_0(\lambda)\rangle$ is the ground state wavefunction, and $\delta\lambda$ a small variation in $\lambda$. 
The peaks of these two quantities serve as approximate indicators of criticality. 
In \cref{fig:App_Rydberg_transition_point}, we show the critical regimes for $16$ atoms on a square lattice ($4\times4$) and on a ladder ($8\times2$), where we used $\delta\lambda=0.002$.

\section{Adiabatic echo protocol with $\mathbb{Z}_3$ and $\mathbb{Z}_4$ symmetry breaking}

In the main text, we presented an analytical argument explaining the robustness of the echo protocol in the presence of $\mathbb{Z}_2$ symmetry breaking, based on the first-order expression for the preparation infidelity given in \cref{eq:infidelity_SM}. For systems with $\mathbb{Z}_n$ symmetry, this expression generalizes to
\begin{equation}
\mathcal{I} \simeq \epsilon^{2} \sum_{k=1}^{n-1} \left| \int_{0}^{T} dt\, V_{k0}(s(t))\, e^{-i \int_{t}^{T} d\tau\, \delta E_k(s(\tau))} \right|^{2},
\end{equation}
where $V_{k0}(s) = \langle E_k(s)|\hat{V}|E_0(s) \rangle$ and $\delta E_k(s) = E_k(s) - E_0(s)$ are, respectively, the matrix elements and energy gaps between the ground state $|E_0(s)\rangle$ and the first $n-1$ excited states $|E_k(s)\rangle$. We assume that $|E_0(s)\rangle$ is even under $\mathbb{Z}_n$, while each of the excited states $|E_k(s)\rangle$ carries a distinct nontrivial $\mathbb{Z}_n$ charge $k = 1, \dots, n-1$. In the ordered phase, where the $\mathbb{Z}_n$ symmetry is spontaneously broken, the corresponding energy gaps $\delta E_k$ become exponentially small in the system size.

Applying the same reasoning as in the $\mathbb{Z}_2$ case to an echo protocol with turning points at $t_1$, $t_2$, and $t_3$, we obtain 
\begin{equation}
\mathcal{I} \simeq \epsilon^2 \sum_{k=1}^{n-1} \left|\, e^{-i \alpha_k} \cdot \int_{t_1}^{t_2} dt\, V_{k0}(s(t)) + \int_{t_3}^{T} dt\, V_{k0}(s(t)) \,\right|^{2},
\end{equation}
where $\alpha_k = \int_{t_2}^{t_3} d\tau\, \delta E_k(s(\tau))$.
For the first-order contributions to cancel, each term in the sum must vanish individually. In general, this is not guaranteed unless both $\delta E_k$ and $V_{k0}(s)$ are independent of the $\mathbb{Z}_n$ charge label $k$. When this condition holds, the argument for destructive interference leading to suppression of infidelity extends directly from $\mathbb{Z}_2$ to $\mathbb{Z}_n$, and the emergence of the echo protocol is similarly justified. Below, we show that, consistent with this expectation, the echo structure arises in two concrete models exhibiting $\mathbb{Z}_3$ and $\mathbb{Z}_4$ symmetry breaking.

\begin{figure}[h!]
    \phantomsubfloat{\label{fig:App_Z3a}}
    \phantomsubfloat{\label{fig:App_Z3b}}
    \phantomsubfloat{\label{fig:App_Z3c}}
    \phantomsubfloat{\label{fig:App_Z3d}}
    \phantomsubfloat{\label{fig:App_Z4a}}
    \phantomsubfloat{\label{fig:App_Z4b}}
    \phantomsubfloat{\label{fig:App_Z4c}}
    \phantomsubfloat{\label{fig:App_Z4d}}
    \includegraphics[width=0.9\columnwidth]{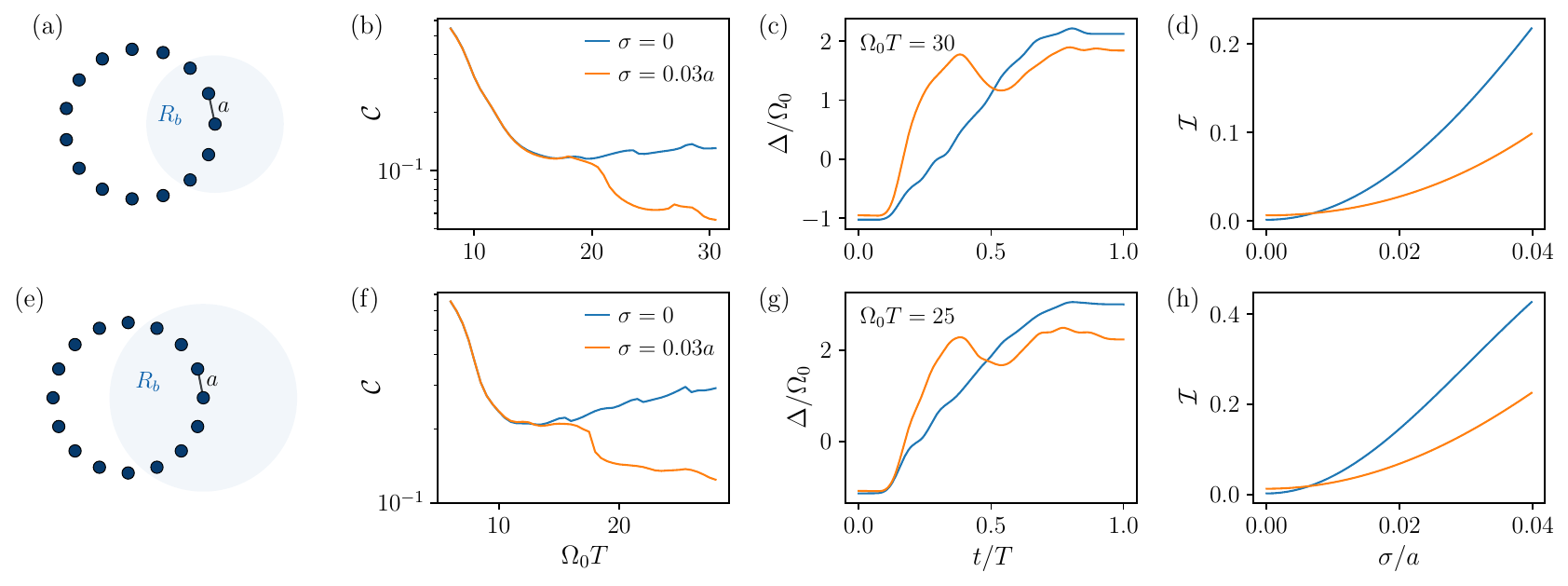}
    \centering
    \vspace*{-3mm}
    \caption{(a,e) $L$ Rydberg atoms on a ring with blockade radius $R_b$: (a) $L=15$ and $R_b=2.2 a$ for $\mathbb{Z}_3$ symmetry breaking; (e) $L=16$ and $R_b = 3 a$ for $\mathbb{Z}_4$ symmetry breaking.  
    (b, f) Optimal cost function in the presence of positional disorder with $\sigma = 0.03 a$ vs total evolution time $\Omega_0 T$. The GRAPE optimization is performed with $\sigma = 0$ (blue line) and $\sigma = 0.03 a $ (orange line). (c, g) Optimal profile for $\Omega_0 T=30$ for $\mathbb{Z}_3$ (c) and $\Omega_0 T=25$ for $\mathbb{Z}_4$ (g). Blue and orange lines correspond to the optimal protocols found for $\sigma = 0$ (standard adiabatic) and $\sigma = 0.03 a$ (adiabatic echo), respectively. (d, h) Preparation infidelity as a function of the disorder strength $\sigma$ for the protocols shown in (c,g).}
    \label{fig:App_Z3Z4}
\end{figure}

We consider Rydberg atoms arranged in a ring geometry, governed by the Hamiltonian in Eq.~(5) of the main text, and study two different blockade radii: $R_b = 2.2a$ and $R_b = 3a$. As in the square lattice and two-leg ladder setups discussed previously, a $\mathbb{Z}_n$ symmetry emerges from the $\mathbb{Z}_L$ translational symmetry of the ring, provided that the number of atoms $L$ is a multiple of $n$ (cf.~\cref{fig:App_Z3a,fig:App_Z4a}). In particular, for $R_b = 2.2a$ with $L$ a multiple of 3, and $R_b = 3a$ with $L$ a multiple of 4, the $\mathbb{Z}_3$ and $\mathbb{Z}_4$ translation symmetries are spontaneously broken at large enough detuning $\Delta$~\cite{Bernien2017}. In this setting, we aim to prepare the $\mathbb{Z}_3$ and $\mathbb{Z}_4$ cat states:
\begin{align}
    \ket{ \mathbb{Z}_3 } & = \frac{1}{\sqrt{3}} \left( \ket{ rgg\cdots rgg } + \ket{ grg\cdots grg } + \ket{ ggr\cdots ggr} \right),  \\
    \ket{ \mathbb{Z}_4 } & = \frac{1}{\sqrt{4}} \left( \ket{rggg\cdots rggg} + \ket{grgg\cdots grgg} + \ket{ ggrg\cdots ggrg } + \ket{ gggr\cdots gggr }  \right),
\end{align}
starting from the product state $\ket{gg\cdots g}$.

Similarly to the Rydberg GHZ preparation of the main text, a natural symmetry-breaking perturbation arises here due to shot-to-shot fluctuations in the atomic positions.
We incorporate these effects in the GRAPE optimization by using as a cost function the preparation fidelity averaged over $N_{\mathrm{s}} = 30$ instances of position displacements, drawn from a gaussian distribution with standard deviation $\sigma$ (cf.~Eq.(4) of the main text).

As in the main text, we set $\Omega(t) = \Omega_0 w(t)$, where $w(t)$ is the cosine-tapered window function plotted in \cref{fig:App_window}, with $\alpha = 0.25$, and use the detuning $\Delta(t)$ as control function. In \cref{fig:App_Z3b,fig:App_Z4b}, we plot the optimal cost function in the presence of positional disorder with $\sigma = 0.03 a$, when the optimization is run with (orange line) and without positional disorder (blue line). This analysis reveals the existence of a threshold time scale $\Omega_0 T^*$, beyond which the optimization that accounts for symmetry-breaking perturbations identifies a protocol that mitigates preparation errors caused by atomic position fluctuations. As shown in \cref{fig:App_Z3c,fig:App_Z4c}, the resulting control profiles exhibit the characteristic features of the adiabatic echo protocol. Finally, in \cref{fig:App_Z3d,fig:App_Z4d}, we demonstrate its robustness by plotting the preparation infidelity as a function of disorder strength $\sigma$, and comparing it to the performance of the standard adiabatic time-optimal protocol.

\section{Robustness of the echo protocol against time-dependent symmetry-breaking noise}\label{app:Time-Dep}

\begin{figure}[h!]
    \phantomsubfloat{\label{fig:App_time_dependent_perta}}
    \phantomsubfloat{\label{fig:App_time_dependent_pertb}}
    \phantomsubfloat{\label{fig:App_time_dependent_pertc}}
    \centering  
    \includegraphics[width=0.8\columnwidth]{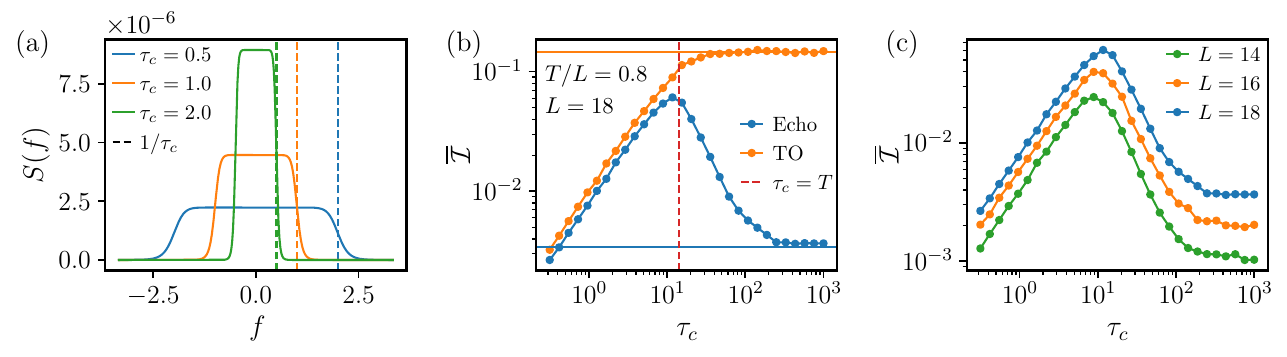}
    \vspace*{-3mm}
    \caption{(a) Power spectral density $S(f)$ of the stochastic longitudinal field $h(t)$ for different noise correlation times $\tau_c$, normalized such that $\int df\, S(f) = \sigma^2$ with $\sigma = 0.003$. In the limit $\tau_c \to 0$, only the zero-frequency component contributes, and $h(t)$ reduces to a time-independent gaussian variable with zero mean and variance $\sigma^2$. (b) GHZ state preparation infidelity for the Ising chain, obtained using the time-optimal (orange) and echo (blue) protocols (cf. Fig.~2b of the main text), as a function of $\tau_c$, averaged over 3000 realizations of the stochastic field $h(t)$. The vertical dashed line marks the preparation time $T$; horizontal solid lines indicate the average infidelities in the strictly static limit. (c) Finite-size scaling of the average infidelity for the echo protocol as a function of $\tau_c$.}
    \label{fig:App_time_dependent_pert}
\end{figure}

In the main text, we derived the echo protocol both analytically, in a general setting, and numerically, in specific models, under the assumption of static symmetry-breaking imperfections. Here, we benchmark the performance of the echo protocol—optimized for the quantum Ising chain (Eq.~(3) of the main text)—in the presence of a time-dependent longitudinal field $h(t)$, and compare it to that of a standard adiabatic time-optimal protocol.
The stochastic field $h(t)$ is defined through its two-time correlation function,
\begin{equation}
\langle h(t) h(0) \rangle  =  \int_{-\infty}^{+\infty} \! df \, e^{i 2 \pi f t } S(f)
=  \int_{-\infty}^{+\infty} \! df \, e^{i 2 \pi f t } \frac{ \mathcal{A}_\sigma }{ 1 + (\tau_c f)^{16} },
\end{equation}
where we pick a power spectral density $S(f)$ that decays sharply for frequencies $f \gtrsim 1/\tau_c$, and the normalization constant $\mathcal{A}_\sigma$ is chosen such that $\int_{-\infty}^{+\infty} S(f) \, df = \sigma^2$ (cf. \cref{fig:App_time_dependent_perta}). We set $\sigma = 0.003$ in what follows. 
The noise is effectively static for $\tau_c \gg T$ and becomes Markovian for $\tau_c \ll T$, with $T$ the preparation time.

We generate time series of the stochastic field using standard techniques from signal processing,  
\begin{equation}
    h(t) = \sum_k \sqrt{2 S(f_k) \, \Delta f} \cos(2\pi f_k t + \phi_k),
\end{equation}
where $\phi_k$ are random phases uniformly distributed in $[0, 2\pi)$ and $\Delta f$ is the frequency resolution.  
We then compute the GHZ preparation infidelity using both the standard adiabatic and echo protocols (with $T/L = 0.8$, as in Fig.~2b of the main text), averaging over $3000$ realizations of $h(t)$. In \cref{fig:App_time_dependent_pertb}, we compare the average infidelity $\mathcal{I}$ of the two protocols as a function of the noise correlation time $\tau_c$.  
In the Markovian regime ($\tau_c \lesssim T$), the echo protocol offers no advantage, since interference effects cannot be exploited to suppress incoherent errors. By contrast, in the quasi-static regime ($\tau_c \gtrsim T$), the echo protocol performs significantly better than the time-optimal protocol, thereby demonstrating its effectiveness against slowly varying symmetry-breaking noise. Note that, in the Markovian limit, the white noise strength goes to $0$ due to the normalization $\int_{-\infty}^{+\infty} S(f) \, df = \sigma^2$, leading to a vanishing infidelity for $\tau_c/T \to 0$.

\end{document}